# Frequency preference in two-dimensional neural models: a linear analysis of the interaction between resonant and amplifying currents

Horacio G. Rotstein (horacio@njit.edu)
Farzan Nadim (farzan@njit.edu)

Department of Mathematical Sciences, New Jersey Institute of Technology
Department of Biological Sciences, New Jersey Institute of Technology, and Rutgers University-Newark

## Abstract

Many neuron types exhibit preferred frequency responses in their voltage amplitude (resonance) or phase shift to subthreshold oscillatory currents, but the effect of biophysical parameters on these properties is not well understood. We propose a general framework to analyze the role of different ionic currents and their interactions in shaping the properties of impedance amplitude and phase in linearized biophysical models and demonstrate this approach in a two-dimensional linear model with two effective conductances $g_L$ and $g_1$. We compute the key attributes of impedance and phase (resonance frequency and amplitude, zero-phase frequency, selectivity, etc.) in the $g_L$-$g_1$ parameter space. Using these attribute diagrams we identify two basic mechanisms for the generation of resonance: an increase in the resonance amplitude as $g_1$ increases while the overall impedance is decreased, and an increase in the maximal impedance, without any change in the input resistance, as the ionic current time constant increases. We use the attribute diagrams to analyze resonance and phase of the linearizations of two biophysical models that include resonant ($I_h$ or slow potassium) and amplifying currents (persistent sodium). In the absence of amplifying currents, the two models behave similarly as the conductances of the resonant currents is increased whereas, with the amplifying current present, the two models have qualitatively opposite responses. This work provides a general method for decoding the effect of biophysical parameters on linear membrane resonance and phase by tracking trajectories, parameterized by the relevant biophysical parameter, in pre-constructed attribute diagrams.

# Introduction

A variety of rhythmic oscillations are observed in different areas of the central nervous system which are known to be critical in cognitive and motor behaviors and are thought to emerge from the coordinated activity of the neurons in the respective networks (Gray 1994; Marder and Calabrese 1996; Wang 2010). Recent work suggests that the frequency of the network oscillations may crucially depend on the intrinsic preferred frequencies of the constituent neurons (Lau and Zochowski 2011; Ledoux and N. 2011; Tohidi and Nadim 2009; Wu et al. 2001). These preferred frequencies arise in different contexts: the ability of a neuron to generate sub-threshold oscillations at a particular frequency, often in response to a DC current input (Dickson and Alonso 1997; Lampl and Yarom 1997; Reboreda et al. 2003; Schmitz et al. 1998); the tendency of a neuron to produce subthreshold membrane potential resonance, a peak in the impedance amplitude in response to an oscillatory input current at a non-zero (resonant) frequency (Hutcheon and Yarom 2000); or membrane potential oscillations with zero phase lag in response to an oscillatory current input at a specific frequency (zero-phase frequency).

The properties of membrane resonance has been investigated in various systems (Dembrow et al. 2010; Gutfreund et al. 1995; Hu et al. 2002; Llinás and Yarom 1986; Narayanan and Johnston 2007; Reboreda et al. 2003; Reinker et al. 2004; Schreiber et al. 2004; Tohidi and Nadim 2009) but the biophysical mechanisms underlying the generation of membrane resonance and its relationship to suprathreshold oscillations are not well understood (Richardson et al. 2003). Membrane resonance results from a combination of low- and high-pass filter mechanisms that also require a negative feedback effect (Hutcheon and Yarom 2000). The passive membrane currents typically act as a low-pass filter, and the so-called resonant voltage-gated currents oppose voltage changes at low frequencies and act as a high-pass filter, thus creating a preferred frequency band. Other voltage-gated ionic currents, labeled amplifying currents, do not produce resonance but generate a positive feedback effect that amplifies voltage changes and hence make existing resonance more pronounced (Hutcheon and Yarom 2000).

Although many studies have examined the impedance amplitude, much less attention has been paid to the phase of impedance which determines the phase of the subthreshold voltage response to oscillatory input currents (Richardson et al. 2003). For linear *RLC* circuits in series, this phase is zero at the resonant frequency. On the other hand, for *RLC* circuits in parallel, such as neural circuits, the zero-phase and the resonant frequencies do not necessarily coincide. We refer to the phase corresponding to the resonant frequency as the resonant phase, to the occurrence of a zero-phase voltage output at a positive input frequency as the zero-phase frequency and refer to the frequency itself as $f_{phase}$. To our knowledge, the mechanisms underlying the occurrence of the zero-phase frequency in biophysical neuronal systems have not been investigated before.

It is clear that the parameters of resonant and amplifying ionic currents shape the impedance and phase profiles but the relationship between these factors is far from intuitive. For instance, as we show in this study, increasing the maximal conductance of a so-called resonant current may lead to an increase or decrease of the maximal impedance value depending on the properties of this ionic current. As such, the characterization of the impedance and phase profiles (and the corresponding subthreshold resonance frequency and zero-phase frequency) in terms of the properties of the ionic current parameters can provide valuable information on how interactions among biophysical properties of neurons may lead to emergent preferred frequencies and, in turn, influence network oscillations and synchrony.

The existence of subthreshold oscillations indicates an intrinsic time scale of the neuron comparable to the subthreshold oscillation period. The resonant frequency, on the other hand, reflects an emergent time scale which is a property of the interaction between the neuron and the input it receives. Intrinsic subthreshold oscillations and subthreshold resonance are related (Lampl and Yarom 1997) but are not equivalent phenomena. Linear models, for instance, may exhibit one but not the other (Richardson et al. 2003). Similarly, emergent time scales may be different from the neuron's intrinsic time scale, and are expected to be more relevant in the communication of subthreshold activity to the spiking and network activity. For instance, with large enough DC current, spikes are more likely to occur at the subthreshold oscillation peaks, whereas, with increased amplitude levels of an oscillatory input, spikes are more likely to occur at the resonant phase and inherit the resonant frequency.

In this paper we investigate the biophysical mechanisms underlying the emergence of the resonance and zero-phase frequencies in two-dimensional models. In particular, we examine the role that different ionic currents and their associated time constants play in determining the resonant and zero-phase frequencies and in shaping other relevant properties of a neuron's voltage response to oscillatory inputs. Our study consists of two parts. First, we investigate the basic mechanisms of generation of resonance and zero-phase frequency in a reduced two-dimensional linear model with two effective dimensionless conductances, a leak conductance $g_L$ and a "resonant" conductance $g_1$ (Richardson et al. 2003). This reduced system represents a linearization of a non-dimensionalized two-variable conductance-based model. We construct diagrams of the basic attributes (resonant and zero-phase frequencies, amplitude, etc.) of the impedance and phase profiles in the $g_L - g_1$ parameter space. Secondly, we then investigate the effect of $g_L$ and $g_1$ on these attributes by exploring changes in their properties along horizontal (constant $g_1$) and vertical (constant $g_L$) lines in the corresponding attribute diagrams. In these reduced linear models, the kinetics of the resonant gating variable is not explicit but included in the effective conductances (Richardson et al. 2003). Additionally, we show that the effect of the corresponding time constants can be examined in the attribute diagrams along oblique lines in the $g_L - g_1$ parameter space.

In the second part of the study we use the attribute diagrams mentioned above to investigate the mechanisms of generation of resonance and zero-phase frequency in two-dimensional linearized conductance-based models. We consider two types of models, each with a resonant current. In the first model, the resonant current is a hyperpolarization-activated inward current $I_h$ (Haas and White 2002; Hutcheon et al. 1996; Schreiber et al. 2004). In the second model, the resonant current is a slow potassium current $I_{Ks}$ activated by depolarization (Gutfreund et al. 1995). In each model we also examine the effect of a fast amplifying current $I_P$ (Gutfreund et al. 1995; Haas and White 2002; Hutcheon et al. 1996; Schreiber et al. 2004) considered here with instantaneous kinetics. The study of the interaction between two distinct resonant currents with slow dynamics leads to models with at least three variables and is outside the scope of this paper. Changes in the biophysical conductances associated with these currents are reflected in changes in the effective conductances $g_L$ and $g_1$. The model nonlinearities are captured by nonlinear trajectories in the $g_L - g_1$ parameter space as one of the biophysical conductances change. We show that, although both $I_h$ and $I_{Ks}$ are resonant currents, the dependence of the resonance attributes on the biophysical conductances, in particular, the mechanisms underlying the amplification of the voltage responses are qualitatively different.

## Methods

### *Conductance-based models*

In this paper we consider biophysical (conductance-based) models of Hodgkin-Huxley type (Hodgkin and Huxley 1952). The current-balance equation is given by

$$C\frac{dV}{dt} = -I_L - \sum_k I_{x_k} + I_{app} + I_{in}(t) \tag{1}$$

where $V$ is the membrane potential (mV), $t$ is time (msec), $C$ is the membrane capacitance (μF/cm$^2$), $I_{app}$ is the applied DC current (μA/cm$^2$), $I_L = G_L(V - E_L)$ is the leak current, and $I_{x_k}$ are generic expressions for ionic currents of the form

$$I_x = G_x x(V - E_x) \tag{2}$$

with maximal conductance $G_x$ (mS/cm$^2$) and reversal potentials $E_x$ (mV). All gating variables $x$ follow a first order differential equation of the form

$$\frac{dx}{dt} = \frac{x_\infty(V) - x}{\tau_x(V)} \tag{3}$$

where $x_\infty(V)$ and $\tau_x(V)$ are, respectively, the voltage-dependent activation/inactivation curves and time constants. The function $I_{in}(t)$ is a time-dependent input current (in µA/cm²). For sinusoidal input currents, we use the following notation

$$I_{in}(t) = A_{in}\sin(\omega t) \quad \text{with} \quad \omega = \frac{2\pi f}{T} \tag{4}$$

Where $T = 1000$ msec and $f$ is in Hz.

The generic ionic currents (Eq. (2)) we consider here are restricted to have a single gating variable $x$ and are linear in $x$. This choice is motivated by the persistent sodium current $I_P$, hyperpolarization-activated inward current $I_h$ and slow M-type potassium current $I_{Ks}$ found in a variety of neurons that exhibit subthreshold resonance (Acker et al. 2003; Rotstein et al. 2006; Schreiber et al. 2004). Ionic currents of this form have been used in a variety of other models (Morris and Lecar 1981; Rinzel and Ermentrout 1998). The methods we use in this paper can be easily adapted for currents with two gating variables and gating variables raised to powers higher than 1.

We focus on two-dimensional models that describe the dynamics of voltage $V$ and a gating variable $x_1$, respectively. Additional currents whose gating variables evolve on a fast time scale (faster than all other variables), such as persistent sodium currents, can be included by using the adiabatic approximation $x_k = x_{k,\infty}(V)$. Here we consider one such additional (generic) current $I_{x_2}$ with $x_2 = x_{2,\infty}(V)$. Additional fast currents can be considered without significantly changing the formalism used here.

*Linearized conductance-based models*

Linearization of conductance-based models around a stable fixed-point $(\bar{V}, \bar{x}_1)$ with $\bar{V}$ below threshold for spike generation is a standard procedure (Ermentrout and Terman 2010). In its simplest form, it consists of substituting the right hand side of Eqs. (1) and (3) by their first order Taylor expansions calculated at the fixed point.

Here we follow Richardson et al. (Richardson et al. 2003) and linearize the autonomous part of system (1)-(3) around the fixed-point $(\bar{V}, \bar{x}_1)$ for the isolated system ($I_{in}(t) = 0$) by defining

$$v = V - \bar{V}, \quad w_1 = \frac{x_1 - \bar{x}_1}{x'_{1,\infty}(\bar{V})} \tag{5}$$

where $\bar{x}_1 = x_{1,\infty}(\bar{V})$. The linearized equations are given by (Richardson et al. 2003)

$$C\frac{dv}{dt} = -g_L v - g_1 w_1 + I_{in}(t)$$
$$\bar{\tau}_1 \frac{dw_1}{dt} = v - w_1 \qquad (6)$$

where the effective leak and ionic conductances and time constants are defined by

$$g_L = G_L + G_1 + G_2 + g_2$$
$$G_1 = G_{x_1} x_{1,\infty}(\bar{V})$$
$$G_2 = G_{x_2} x_{2,\infty}(\bar{V}) \qquad (7)$$
$$g_2 = G_{x_2} x'_{2,\infty}(\bar{V})(\bar{V} - E_{x_2})$$

$$g_1 = G_{x_1} x'_{1,\infty}(\bar{V})(\bar{V} - E_{x_1}), \quad \bar{\tau}_1 = \tau_{x_1}(\bar{V}) \qquad (8)$$

respectively. Note that the effective leak conductance $g_L$ includes components from both the biophysical leak conductance $G_L$ and the voltage-dependent ionic conductances: the slow ionic currents contributes a single term $G_1$ whereas the fast current $I_{x_2}$ contributes two terms, $G_2$ and $g_2$.

The effective ionic conductance $g_1$ can be either positive or negative. The sign of $g_1$ determines whether the associated gating variable is resonant ($g_1 > 0$) or amplifying ($g_1 < 0$) (Hutcheon and Yarom 2000; Richardson et al. 2003). In our discussion, we will limit the values of $g_1$ to the positive range which produces resonance. Note that $g_2$, the last term of $g_L$, may also be positive or negative, depending on whether $\bar{V} > E_{x_2}$, and is of the same form as $g_1$. Thus, if $g_2$ is negative, the fast current is amplifying.

Note that the gating variable $w_1$ in (5) has units of voltage. Note also that the linearization process described above is not restricted to ionic currents of the form (2) and can be applied to models with an arbitrary number of ionic currents and any voltage-dependent time constant (Richardson et al. 2003).

System (6) can be rescaled in order to reduce the number of parameters (from four to two) that effectively govern its dynamics with no loss of information. We follow (Richardson et al. 2003) and define the following dimensionless time and parameters

$$\hat{t} = \frac{t}{\bar{\tau}_1}, \quad \gamma_L = \frac{g_L \bar{\tau}_1}{C}, \quad \gamma_1 = \frac{g_1 \bar{\tau}_1}{C}. \qquad (9)$$

Substituting into (6) we obtain

$$\frac{dv}{d\hat{t}} = -\gamma_L v - \gamma_1 w_1 + \hat{I}_{in}(\hat{t})$$
$$\frac{dw_1}{d\hat{t}} = v - w_1 \qquad (10)$$

where

$$\hat{I}_{in}(t) = \hat{A}_{in} \sin(2\pi f \hat{t} / \hat{T}) \text{ with } \hat{A}_{in} = \frac{A_{in} \overline{\tau}_1}{C} \qquad (11)$$

with $\hat{T} = T / \overline{\tau}_1 = 1000 / \overline{\tau}_1$. Note that $[f] = $ Hz, and $[v] = [w_1] = V$. We refer to $\gamma_L$ and $\gamma_1$ as the dimensionless effective conductances.

*Impedance and impedance-like functions*

The voltage response of a linear system receiving sinusoidal current inputs of the form (4) is given by

$$V_{out}(t; f) = A_{out}(f) \sin(\omega t - \phi(f)) \qquad (12)$$

where $\phi(f)$ is the phase offset (time difference between the peaks of the input current and the output voltage normalized by $2\pi$) and $\omega$ is given by the second equation in (4). Linear systems exhibit subthreshold (amplitude) resonance, if there is a peak in ratio of the output voltage and input current amplitudes:

$$Z(f) = \frac{A_{out}(f)}{A_{in}} \qquad (13)$$

at some positive (resonant) frequency $f_{res}$, and zero-phase frequency, if the phase offset $\phi(f)$ vanishes at some positive frequency $f_{phase}$. We calculate the phase as the time difference between the output voltage and input current peaks normalized by the oscillation period. For simplicity, we will refer to the amplitude of impedance as impedance.

Figs. 1A and 1B show two representative graphs for the impedance profile of a low-pass filter RC model (panel A) and a band-pass filter model (panel B) exhibiting resonance at $f = f_{res}$. We will use four parameters to characterize impedance profiles:

(i) the resonant frequency $f_{res}$,
(ii) the maximum impedance $Z_{max} = Z(f_{res})$,
(iii) the resonance amplitude or power $Q_Z = Z_{max} - Z(0)$, and
(iv) the characteristic right-band-width $\Lambda_{1/2}$ defined as the range of values of $f \geq f_{res}$ such that $Z(f) \geq Z_{max} / 2$.

$\Lambda_{1/2}$ provides information about the selectivity of the neuron to input frequencies. The smaller the value of $\Lambda_{1/2}$, the higher the selectivity, i.e., the impedance profile is peakier around the resonant frequency $f_{res}$. The choice of ½ is arbitrary and other values can be used. The restriction $f \geq f_{res}$ is useful for comparison between cases where $Z(0) > Z_{max} / 2$ in one or more of the impedance profiles. In these cases, a "full" bandwidth will not be representative of the difference between impedance profiles. For a low-pass filter RC cell as in Fig. 1A, $f_{res} = 0$ and $Q_Z = 0$.

Figs. 1C and 1D show two representative graphs for the phase ($\phi$). They both approach $\pi/2$ for large values of $f$. The phase in panel C is always increasing and positive; the voltage response is always delayed by at most one quarter of a cycle. The phase in panel D, on the other hand, is negative for $f < f_{phase}$ and positive for larger values of $f$. The negative phase can be interpreted as the response leading the input. For $f = f_{phase}$, voltage response and input are in phase, so $f_{phase}$ represents the zero-phase frequency. We characterize the phase profile with an additional parameter $\phi_{min}$ measuring the minimum phase.

In this paper we consider sinusoidal current inputs of the form (4) where $A_{in}$ is a non-negative constant and the frequency $f$ is measured in number of oscillations per 1000 units of time. For dimensional consistency with most experimental measurements, time $t$ is expressed in milliseconds and $f$ is measured in Hz. We use the same terminology for sinusoidal inputs with dimensionless time.

## Results

### 1. Generation of resonance and zero-phase frequency in two-dimensional linear models

In the first part of the Results we consider the 2D linear system (6) and sinusoidal current inputs $I_{in}(t)$ of the form (4) with constant values of $A_{in}$. Because the system is linear, without loss of generality, we assume $A_{in} = 1$. For simplicity in the notation, we drop the "bar" from $\tau_1$ and the "hat" from $t$ and other parameters in (6) and (10). The eigenvalues, impedance, and resonant frequency for system (6) can be calculated from Eqs. (18), (21) and (22) in Appendix A by substituting $a = -g_L C^{-1}$, $b = -g_1 C^{-1}$, $c = \tau_1^{-1}$, and $d = -\tau_1^{-1}$. Alternatively, one can use the rescaled system (10) and compute the eigenvalues, impedance, and resonant frequency ($\hat{r}_{1,2}$, $\hat{Z}$ and $\hat{\omega}_{res}$ respectively) by substituting $a = -\gamma_L$, $b = -\gamma_1$, $c = 1$ and $d = -1$ in Eqs. (18), (21) and (22) of Appendix A. This information can then be used in the two-dimensional parameter space of effective conductances to investigate the effects of the biophysical parameters $g_L$, $g_1$, $\tau_1$, and $C$ using

$$r_{1,2} = \frac{\hat{r}_{1,2}}{\tau_1}, \ Z(\omega) = \frac{\tau_1}{C}\hat{Z}(\tau_1 \omega), \text{ and } f_{res} = \frac{\hat{\omega}_{res}}{\tau_1}\frac{1000}{2\pi}. \tag{14}$$

For $C=1$ and $\tau_1 =1$, $g_L = \gamma_L$ and $g_1 = \gamma_1$ the two systems (6) and (10) are equivalent. Without loss of generality, we let $C=1$ and only consider the effects of changing $\tau_1$. Changes in $\tau_1$ affect the magnitude of the eigenvalues but not their sign. These changes, therefore, do not affect the stability properties of the autonomous system but do affect, for instance, the natural frequency $f_{nat}$ (when the autonomous system exhibits damped oscillations). Changes in $\tau_1$ also affect the shape of the impedance function and the resonant frequency as will be discussed.

In Fig. 2A1 (magnified in 2A2) we show the stability diagram for the 2D linear system (6) with $\tau_1 =1$. This diagram is similar to that presented in (Richardson et al. 2003). The blue curves separate regions in parameter space with different stability properties or distinct fixed point types (focus, node or saddle). The system is stable in the regions marked with blue symbols and unstable in regions marked with red. The green curve, referred to as the resonance curve, separates the regions in parameter space for which the system exhibit resonance (shaded green) and exhibit no resonance. Note that the regions in parameter space where the system exhibits intrinsic oscillations and resonance do not coincide (Richardson et al. 2003). In particular, resonance can occur in the absence of intrinsic oscillations where the fixed point is a node.

Representative impedance profiles show a variety of possible resonant behaviors (Figs. 2B to 2E). For $g_1 = 0$, the first equation in the 2D linear system is independent of $w_1$; the system has no resonance and simply describes a low-pass filter (Fig. 2B). As $g_L$ increases, $Z_{max}$ ($= Z(0)$) decreases (and $\Lambda_{1/2}$ increases linearly; not shown). In this case, the phase $\phi$ shows a delayed response for all values of $g_L$ but this delay decreases as $g_L$ increases (not shown). The expression of resonance in this system occurs when $g_1 > 0$. For any fixed positive value of $g_1$, increasing the value of $g_L$ results in an increase in the value of the resonance frequency $f_{res}$ and a concomitant decrease in $Z_{max}$ (Figs. 2C and 2D). Note, however, that resonance can occur with $g_L = 0$ (Fig. 2E).

### *An increase in $g_1$ generates resonance by an unbalanced decrease in $Z_{max}$ and $Z(0)$*

In order for the 2D linear system to exhibit resonance, the value of $g_1$ has to be in the resonant region above the resonance curve in Fig. 2A. For small enough values of $g_1$, the system acts as a low-pass filter similar to the $g_1 = 0$ case described above. For resonance to emerge, a difference between $Z_{max}$ and $Z(0)$ (hence, a positive value of $Q_Z = Z_{max} - Z(0)$) has to be generated as $g_1$ crosses the resonance curve. Figs. 3A and 3B show that below the resonance curve both the $Z_{max}$ (and $Z(0)$) level curves are almost linear and parallel to the line $g_1 + g_L = 1$, and they also nearly coincide (e.g., see Fig. 3C for $\gamma_1 < 0.5$). Above the resonance curve, the $Z_{max}$ level sets are no longer linear while the $Z(0)$ level sets remain linear. Thus, as $g_1$ increases along vertical lines in $g_L$-$g_1$ parameter space, resonance emerges because $Z(0)$ decreases faster than $Z_{max}$ (Fig. 3C). This difference $Q_Z$ is more attenuated for larger values of $g_L$ (not shown). As expected, when $g_L$ increases, both $Z_{max}$ and $Z(0)$ decrease as $1/g_L$.

The resonance line is marked by a peak in the half-band-width $\Lambda_{1/2}$ in the $g_1$-$g_L$ parameter space as $g_1$ increases (Figs. 3E and 3F). As $g_1$ increases further, $\Lambda_{1/2}$ settles into an almost constant value. The selectivity of resonance is therefore independent of $g_1$ for large enough values of $g_1$ (Fig. 3F). In contrast, $\Lambda_{1/2}$ increases monotonically with $g_L$ (Fig. 3G).

*An increase in $g_L$ generates low amplitude resonance*

Resonance occurs for all values of $g_L$ within the stability region for the underlying autonomous system, provided $g_1$ is large enough to be above the resonance curve (Fig. 2A, green region). For smaller values of $g_1$ (as in Fig. 2C), resonance can be generated by increasing $g_L$ along horizontal lines that cross the resonance curve in Fig. 2A. Along these lines, the resonance frequency increases with $g_L$, and both $Z_{max}$ and $Z(0)$ decrease but at a slightly different pace (Fig. 3D), thus generating resonance. The resonance generated by this mechanism is much weaker as compared to the one described above for increasing values of $g_1$ in the sense that $Q_Z$ is much smaller. This is because both $Z_{max}$ and $Z(0)$ change along almost parallel curves (e.g., Fig. 3D).

*A decrease in $g_L$ amplifies the voltage response and decreases $f_{res}$*

A decrease in the effective leak conductance $g_L$ (Eq. (7)) can be caused by a decrease in the biophysical leak conductance $G_L$, a decrease in the ionic conductances $G_1$ and $G_2$, or an increase in the fast ionic conductance $g_2$ when the associated gating variable is amplifying ($g_2 < 0$; seen Methods). In the latter case there is a competition between the relative magnitudes of the last two terms in Eq. (7). In some cases, as with the biophysical model we discuss later in this paper, the effective leak conductance $g_L$ can in fact be negative.

Figs. 3A and 3B show that, as $g_L$ decreases, both $Z_{max}$ and $Z(0)$ increase, and so does $Q_Z$ due to the different rates at which $Z_{max}$ and $Z(0)$ change in parameter space (Fig. 3D; see also Fig. 2D). The amplification of the voltage response is more dramatic for values of $g_L$ close to the boundary between the stability and instability regions, and is more pronounced for larger values of $g_1$. In addition to these changes, $\Lambda_{1/2}$ decreases with $g_L$; i.e., as $g_L$ decreases, the impedance profile becomes narrower and the selectivity increases (Figs. 3E and 3G). For values of $g_1$ close to the resonance curve, for which $Z_{max}$ and $Z(0)$ are not significantly different, resonance can be abolished for larger values of $g_L$ (see Fig. 2D).

Together, these results show that a decrease in $g_L$ causes an amplification of resonance, marked by an increase in the resonance amplitude $Q_Z$ and an increase in the selectivity (decrease in $\Lambda_{1/2}$), accompanied by a decrease in $f_{res}$. An increase in $g_L$, therefore, causes an attenuation of resonance accompanied by an increase in $f_{res}$.

### *Resonance and subthreshold oscillations*

To explore the distinction between resonance and subthreshold oscillations in the 2D linear model (i.e., the presence of a focus fixed point), we plotted the resonance and natural oscillation frequencies in the $g_L$-$g_1$ parameter space (Fig. 4). A frequency of 0 indicates lack of oscillation or resonance, as indicated. We found that both $f_{nat}$ and $f_{res}$ increase with $g_1$ but do not generally coincide (see, for example, Fig. 4C). As $g_L$ is increased, both $f_{nat}$ and $f_{res}$ increase for small $g_L$ values but only $f_{res}$ continues to increase with large $g_L$ values whereas $f_{nat}$ decreases back to 0 in this range (Fig. 4D). Note that if $g_1$ is large (e.g., $g_1 = 1$ here), $f_{res}$ and $f_{nat}$ almost coincide during the increasing part of $f_{nat}$ as $g_L$ increases (not shown).

### *An increase in $g_1$ generates zero-phase frequency with $f_{phase} \neq f_{res}$*

We examined how the response phase of the membrane potential shifts with respect to the sinusoidal input current for the 2D linear system (6) with $\tau_1 = 1$ and under what conditions the system exhibits zero-phase frequency. For $g_1 = 0$ (low-pass filter), the phase $\phi$ is always positive with $\phi_{min} = \phi(0) = 0$ and the system does not exhibit zero-phase frequency. Fig. 5A shows that this persists for values of $g_1 < 1$ for which the 2D system exhibits resonance but not zero-phase frequency. For $g_1 > 1$, a negative frequency band of length $f_{phase}$ emerges (see also Fig. 5C). This band starts at $f$ just above zero and its width also determines the zero-phase frequency ($= f_{phase}$). The value of $f_{phase}$ increases with $g_1$ but is independent of $g_L$. It should be noted that $f_{phase}$ is different from the resonant frequency $f_{res}$, and the two variables have different monotonic dependencies on $g_1$. Note also that the minimum phase $\phi_{min}$ increases with $g_1$, and therefore with $f_{phase}$, but decreases with $g_L$ (Fig. 5B).

### *The effect of the time constant $\tau_1$ on resonance*

We now consider the effect of the time constant $\tau_1$ of the gating variable $w_1$ (Eq. (6)). Although Figs. 2 to 5 were produced with $\tau_1 = 1$, the effects of changes in $\tau_1$ can be investigated using these figures and Eqs. (14). When $\tau_1 \neq 1$, $\gamma_L \neq g_L$ and $\gamma_1 \neq g_1$. For fixed values of $g_L$ and $g_1$, changes in $\tau_1$ cause $\gamma_L$ and $\gamma_1$ to move in parameter space along lines (parametrized by $\tau_1$) with slopes $\gamma_1/\gamma_L = g_1/g_L$ as shown in Fig. 6A. Each of these lines converges to the origin as $\tau_1$ approaches zero, and its slope has the same sign as $g_L$ (in the region of resonance for which $g_1 > 0$). Fig. 6A shows that, for small enough values of $\tau_1$, the system exhibits neither intrinsic (damped) oscillations nor resonance, reflecting the fact that both phenomena depend upon a (relatively) slow negative feedback.

As $\tau_1$ increases, the system exhibits first intrinsic oscillations with no resonance for small $\tau_1$ and then, for larger values of $\tau_1$, both intrinsic oscillations and resonance (Figs. 6B1-3; most clearly seen in 6B2). Depending on the slope $\gamma_1/\gamma_L$, intrinsic oscillations may disappear as $\tau_1$ increases further. Note that for $g_L < 0$, relatively small increases in $\tau_1$ causes the system to reach the instability region in parameter space.

An increase in $\tau_1$ amplifies the voltage response and this amplification is marked by an increase in $Q_z$, $Z_{max}$ (Fig. 6C1-3) and a decrease in $\Lambda_{1/2}$ (i.e., increase in selectivity). Note that the amplification of the voltage response is stronger, the larger the difference between $g_1$ and $g_L$. The mechanism of generation of resonance as $\tau_1$ increases is different from the mechanism described above for increasing values of $g_1$ (and constant $\tau_1$). Direct substitution of the model parameters ($g_L$, $g_1$ and $\tau_1$) into Eq. (21) in Appendix A yields $Z(0) = (g_L + g_1)^{-1}$. In other words, changes in $\tau_1$ do not affect the value of $Z(0)$ and hence $Q_Z = Z_{max}$ for all values of $\tau_1$. As such, an increase in the time constant $\tau_1$ generates resonance by an increase in $Z_{max}$ with no changes in $Z(0)$ (Fig. 6C1-3).

Increasing the time constant $\tau_1$ causes the natural and resonant frequencies to increase for small values of $\tau_1$ and to decrease for larger values. Information about the effect of $\tau_1$ on the resonant and natural frequencies can, in principle, be extracted from Figs. 6B and 6C and the expressions for $r_{1,2}$ and $f_{res}$ in Eqs. (14). However, in most cases, especially when $f_{nat}$ or $f_{res}$ increase along lines in parameter space, the net effect of increasing values of $\tau_1$ is not readily apparent from the graphs because both the numerator and denominator are increasing functions of $\tau_1$. In these cases, it is more useful to analyze the expressions $f_{nat}$ and $f_{res}$ given by Eqs. (19) and (22) in Appendix A. Substituting $a = -g_L$, $b = -g_1$, $c = \tau_1^{-1}$, and $d = -\tau_1^{-1}$ yields

$$f_{nat} = \frac{1000}{2\pi} \frac{\sqrt{4g_1\tau_1 - (1 - g_L\tau_1)^2}}{2\tau_1}$$

$$f_{res} = \frac{1000}{2\pi} \frac{\sqrt{-1 + \sqrt{g_1^2\tau_1^2 + 2g_L g_1\tau_1^2 + 2g_1\tau_1}}}{\tau_1}.$$

(15)

Equation (15) gives the imaginary part of the eigenvalues (18) and is valid for calculating $f_{nat}$ so long as the eigenvalues are complex. As such, $f_{nat}$ is zero for the minimum value of $\tau_1$ for which it is defined, reaches a maximum at $\tau_1 = (2g_1 + g_L)^{-1}$ (provided this value is positive) and is zero again for large enough $\tau_1$ when the right hand side of (15) fails to be a real number. $f_{res}$ is zero for the minimum value of $\tau_1$ for which it is defined ($\tau_1 = (-g_1 + \sqrt{2g_1^2 + 2g_L g_1})/(g_1^2 + 2g_L g_1)$), it is positive for larger values of $\tau_1$ (e.g., $\tau_1 = g_1^{-1}$), and approaches zero as $\tau_1$ approaches infinity (Figs. 6B1-3).

The non-monotonic effect of $\tau_1$ on $f_{res}$ is particularly clear in the impedance profile shown in Fig. 6C2. Note, however, that increasing $\tau_1$ always results in an increase of the peak impedance value as seen in Fig. 6C1-3.

We now examine the effect of the time constant $\tau_1$ on phase. Increasing $\tau_1$ generates zero-phase frequency with $f_{phase} \neq f_{res}$. As before, it is useful to examine the expression for $\phi$ given by Eq. (23) in Appendix A. The zero-phase frequency is obtained when the denominator in Eq. (23) vanishes. Substituting $a = -g_L$, $b = -g_1$, $c = \tau_1^{-1}$, and $d = -\tau_1^{-1}$ yields

$$f_{phase} = \frac{1000}{2\pi} \frac{\sqrt{g_1\tau_1 - 1}}{\tau_1}.$$

(16)

Zero-phase frequency occurs for values of time constant $\tau_1$ above $g_1^{-1}$ as illustrated in Fig. 6D. Figs. 6D and 6C also illustrate that the critical value of $\tau_1$ above which zero-phase frequency occurs is different from the critical value above which resonance occurs. From Eq. (16), the zero-phase frequency $f_{phase}$ increases with $\tau_1$ for $\tau_1 < 2g_1^{-1}$ and decreases for larger values of $\tau_1$. A comparison between Eq. (16) and the second equation in (15) shows that in general $f_{phase} \neq f_{res}$.

## 2. Generation of resonance and zero-phase frequency in linearized biophysical two-dimensional models

The study of the roles of the effective conductances $g_1$ and $g_L$ (or their dimensionless versions $\gamma_1$ and $\gamma_L$) on the generation of both resonance and zero-phase frequency using the diagrams in Figs. 2 to 7 provides a mechanistic insight into these phenomena. However, it does not produce a useful explanation of the mechanisms underlying resonance and zero-phase frequency in realistic biophysical models. In this section we will use the description of the dependence of resonance and zero-phase frequency in the 2D linear system, developed thus far, to explain how linear approximations to biophysical models can be mapped onto our framework of the linear systems and how biophysical ionic currents shape preferred frequencies for impedance and phase.

In the previous sections we kept one parameter fixed and varied another. Changes in either $g_L$ or $g_1$ resulted in trajectories in the $g_L - g_1$ parameter space that moved along horizontal and vertical lines (Figs. 3-5). Additionally, changes in $\tau_1$ generated oblique linear trajectories (Fig. 6). In contrast to these parameters, changes in the biophysical conductances may lead to nonlinear trajectories in the $g_L - g_1$ parameter space. While changes in the biophysical leak ($G_L$) and fast conductance ($G_{x_2}$) affect only the effective leak conductance $g_L$ (Eq. (7)), changes in the biophysical conductances with slower time scales ($G_{x_1}$) affect both $g_L$ and $g_1$.

In this section we investigate the effects of changes in the resonant and amplifying currents on the mechanisms of generation of resonance and zero-phase frequency in two prototypical biophysical models $I_h + I_p$ and $I_{Ks} + I_p$. The governing equations are given in Appendix B. The nonspecific cation $h$ current $I_h$ is a hyperpolarization-activated inward current whereas the slow non-inactivating potassium current (e.g., the M current) $I_{Ks}$ is a depolarization-activated outward current and, as such, both $I_h$ and $I_{Ks}$ are resonant currents (Hutcheon and Yarom 2000). The depolarization-activated current persistent sodium current $I_P$ is an amplifying current (Hutcheon and Yarom 2000). The fact that $I_h$ and $I_{Ks}$ activate by moving the membrane potential in different directions (hyper- or depolarizing) results in qualitatively different effects on resonance and zero-phase frequency properties, and their dependence on the biophysical conductances. These two models were chosen as representative models that are known to produce resonance but the effects of these ionic currents can be replaced by other currents with similar biophysical properties. The inward rectifying potassium currents can also act as amplifying currents; however, these currents are typically active in ranges of voltage below values where subthreshold resonance is observed and we do not examine their effect.

An increase in the level of $I_p$ generates strong nonlinearities in the $I_h + I_p$ and $I_{Ks} + I_p$ models as seen in the nullclines (Fig. 7). In the absence of $I_p$ the nullclines are only weakly nonlinear; i.e., the linearized nullclines around the fixed-point $\bar{V}$ (the intersection between nullclines) provide a good approximation to the nonlinear nullclines. This is true in both models for a large range of values of $\bar{V}$. An increase in the amount of $I_p$ generates a strong nonlinearity (of parabolic type) in both models (Figs. 7A2 and 7B2). If the fixed-point $\bar{V}$ is near the "knee" of the $V$-nullcline, the linearized nullclines do not provide a good approximation to the nonlinear ones, and cannot be expected to capture the dynamics of the nonlinear system. For values of $G_h$ and $G_q$ for which $\bar{V}$ lies further away from the knee, however, along the left branch of the corresponding parabolic-like $V$-nullclines, the nullclines are only weakly nonlinear (not shown). Qualitatively, the phase plane diagrams for the two models are almost mirror images of each other (Fig. 7A vs. 7B). Although an increase in $G_h$ causes $\bar{V}$ to increase while an increase in $G_q$ causes $\bar{V}$ to decrease, both phase planes respond in a similar way to changes in $G_p$.

*Nonlinear effects generated by increases in the level of $I_p$ are captured by the impedance profiles*

Changes in the biophysical ionic conductances ($G_h$, $G_q$, and $G_p$) cause changes in the effective conductances $g_L$ and $g_1$ through Eqs. (7) and (8), and generate trajectories in $\gamma_L$-$\gamma_1$ parameter space (Figs. 8A and 9A). The shapes of the trajectories in $\gamma_L$-$\gamma_1$ parameter space depend not only on the biophysical conductances but also on the value of the fixed-point $\bar{V}$ which, in turn, depends on the shape of the $V$-nullclines. Note that, for linear systems, the location of the fixed-point has no effect on the resonance and zero-phase frequency properties. The nonlinearities of the $V$-nullclines are translated to the nonlinear trajectories in $\gamma_L$-$\gamma_1$ parameter space through $\bar{V}$. If changes in $G_h$, $G_p$ and $G_q$ had no effect on the location of $\bar{V}$, then an increase in $G_p$ would still cause a decrease in the effective conductance $g_L$, and an increase in either $G_h$ or $G_q$ would cause an increase in both effective conductances $g_L$ and $g_1$. In this case, changes in $G_h$ and $G_q$ would have generated oblique linear trajectories in $\gamma_L$-$\gamma_1$ parameter space. If the changes in $\bar{V}$ (as a result of changes in the biophysical conductances) are small, as it occurs when the nullclines are weakly nonlinear, then one would expect trajectories in parameter space to be weakly nonlinear as well, as is the case for the solid trajectories for to $G_p = 0$ (Figs. 8A and 9A). For larger values of $G_p$ for which the nullclines are strongly nonlinear, the corresponding trajectories are also strongly nonlinear (dashed trajectories for $G_p = 0.5$ in Figs. 8A and 9A).

### Changes in the levels of the resonant currents $I_h$ and $I_{Ks}$ in the presence of $I_p$ have qualitatively different effects on the resonance and zero-phase frequency

Resonant currents, or resonant gating variables, are usually classified into one single group (Hutcheon and Yarom 2000; Richardson et al. 2003). In the absence of $I_p$, changes in $G_h$ and $G_q$ generate weakly linear, almost oblique trajectories with positive slope in the corresponding $\gamma_L$-$\gamma_1$ parameter spaces (solid gray lines in Figs. 8A and 9A). In this case, changes in $G_h$ and $G_q$ have similar effect on the resonance and zero-phase frequency properties of the corresponding models.

In contrast, when $I_p$ is present, changes in $G_h$ and $G_q$ affect the shape of the corresponding trajectories in almost opposite ways (Figs. 8A and 9A, dashed curves for $G_p = 0.5$) reflecting the different biophysical roles these two currents play. In the $I_h + I_p$ model, in particular, the direction of motion of the trajectories for $G_p = 0.5$ and $G_p = 0$ are opposite as $G_h$ increases and, for large enough values of $\tau_1$, trajectories cross the line $\gamma_L = 0$ leading to a strong amplification of the voltage response (Fig. 8B2). This is in contrast to the $I_{Ks} + I_p$ model where an increase in $G_q$ leads to a strong de-amplification of the voltage response (Fig. 9B2). The differences between the trajectories in parameter space for the two models indicates that the dependence of their resonant properties on $G_h$ or $G_q$ are different. In particular,

i. the voltage response is amplified for the $I_h + I_p$ model and attenuated for the $I_{Ks} + I_p$ model,
ii. the selectivity increases ($\Lambda_{1/2}$ decreases) for the $I_h + I_p$ model and decreases ($\Lambda_{1/2}$ increases) for the $I_{Ks} + I_p$ model, and
iii. $\phi_{min}$ increases in absolute value for the $I_h + I_p$ model and decreases for the $I_{Ks} + I_p$ model.

An increase in $G_h$ reduces the voltage response for $G_p = 0$ (Fig. 8B1) but amplifies the voltage response for $G_p = 0.5$ (Fig. 8B2). This amplification is much stronger for larger values of $I_{app}$ (not shown). In contrast, an increase in $G_q$ reduces the voltage response for all values of $G_p$ (Figs. 9B1 and 9B2). The amplification of the voltage response is still larger for larger values of $I_{app}$ but this amplification is stronger for lower values of $G_q$ (not shown).

In the $I_h + I_p$ model both $f_{res}$ and $f_{phase}$ increase with increasing values of $G_h$ but $f_{res}$ decreases for high values of both $G_h$ and $G_p$. For $G_p = 0$, both $f_{res}$ and $f_{phase}$ increase as $G_h$ increases (Fig. 10A1; also see Figs. 8B1 and 8C1), and they do so almost linearly once resonance has been generated. In contrast, for $G_p = 0.5$, while $f_{phase}$ increases with increasing values of $G_h$, $f_{res}$ first increases and then decreases (Fig. 10A1). These values of $G_h$ and $G_p$ correspond to a fixed-point located near the knee of the $V$-nullcline (Fig. 7A2). The switch in the monotonic behavior of $f_{res}$ can therefore be attributed to a nonlinear effect.

In the $I_{Ks} + I_p$ model $f_{res}$ and $f_{phase}$ increase with increasing values of $G_q$ (Fig 10B1; also see Figs. 9B1 and 9C1). This is similar to the increase of $f_{res}$ and $f_{phase}$ with $G_h$ in the $I_h + I_p$ model (Fig. 10A1). Note however that, in contrast to the $I_h + I_p$ model, $f_{phase}$ is larger for $G_p = 0.5$ than for $G_p = 0$, and $f_{res}$ is almost insensitive to changes in $G_p$ (Fig. 10B1). For $G_p = 0$ neither model show intrinsic oscillatory activity, but oscillatory activity can emerge for $G_p > 0$ and with $G_h$ above a certain threshold (Fig. 10A1) or $G_q$ below a certain threshold (Fig. 10B1; also see Figs. 10A2 and 10B2). Interestingly, as $G_h$ increases, $f_{res}$ decreases and $f_{nat}$ increases and these frequencies approach the same value (Fig. 10A1). Similarly, the values of $f_{res}$ and $f_{nat}$ approach the same value but, in contrast to the $I_h + I_p$ model, as $G_q$ decreases ($I_{Ks} + I_p$ model; Fig. 10B1). The fact that the nullclines of the two models are qualitatively mirror images of each other suggests that for these parameter values both models might have a common underlying mechanism for the generation of both intrinsic oscillations and resonance.

Similarly, in the $I_h + I_p$ and $I_{Ks} + I_p$ models $Z_{max}$ and $Q_Z$ show opposite monotonic behaviors as $G_h$ and $G_q$ increase for $G_p > 0$. Even though, in both models, when $G_p = 0$, $Z_{max}$ and $Q_Z$ are small and almost insensitive to changes in $G_h$ and $G_q$ (Figs. 10A3 and 10B3), when $G_p > 0$, these models exhibit opposite behaviors as the corresponding resonant conductance increases: in the $I_h + I_p$ model resonance is amplified by increasing $G_h$ while in the $I_{Ks} + I_p$ model resonance is attenuated by increasing $G_q$ (Figs. 10A3 and 10B3). Note that the largest values of both $Z_{max}$ and $Q_Z$ occur in the strongly nonlinear regions in parameter space referred to above. The increase in $Q_Z$ and decrease in $Z_{max}$ as $G_h$ and $G_q$ demonstrate that, in both models, resonance is created by a combined decrease in both $Z_{max}$ and $Z(0)$ as explained in the previous section for the reduced models.

We now turn to the effect of $G_p$ in the two models. $f_{res}$ has a similar monotonic behavior as $G_p$ changes in the two models but $f_{phase}$ has a different monotonic behavior as $G_p$ changes in the two models. Figs. 10A2 and 10B2 show the behavior of both $f_{res}$ and $f_{phase}$ as $G_p$ increases. In both models, $f_{res}$ decreases with increasing values of $G_p$. This change is more pronounced in the $I_h + I_p$ model than in the $I_{Ks} + I_p$ model. The dependence of $f_{phase}$ with $G_p$ is different in the two models: $f_{phase}$ slightly decreases in the former and increases in the latter. Interestingly, in spite of the different monotonic behavior between the two models, the difference between $f_{res}$ and $f_{phase}$ decreases as $G_p$ increases, and both characteristic frequencies become closer in the nonlinear region in parameter space with a high amplifying effect (Figs. 10A2 and 10B2). As mentioned above, neither model shows intrinsic oscillations for low values of $G_p$ but oscillations arise as $G_p$ increases (Figs. 10A2 and 10B2).

The amplification of the voltage response due to increasing values of $G_p$ dramatically increases in the nonlinear region in parameter space in the two models. This is seen in Figs. 10A4 and 10B4 (see also Figs. 8 and 9) where both $Z_{max}$ and $Q_Z$ increase for large values of $G_p$ as either $G_h$ increases) or $G_q$ decreases. This amplification does not occur for $G_p = 0$ or with $G_p = 0.5$ together with lower values of $G_h$ or larger values of $G_q$, suggesting that it results from the strong nonlinearities present in both models for large values of $G_p$.

### $f_{res}$ and $f_{phase}$ decrease with increasing values of the time constant $\tau_1$ in the two models

For the $I_h + I_p$ model (Fig. 11A1), $f_{res}$ and $f_{phase}$ are both decreasing functions of $\tau_1$, and are both larger for $G_p = 0$ than for $G_p > 0$. For $G_p > 0$, the difference between $f_{res}$ and $f_{phase}$ decreases and they approach identical values. For the $I_{Ks} + I_p$ model (Fig. 11B2), $f_{phase}$ first increases with $\tau_1$ for $G_p = 0$ and then decreases. For $G_p = 0.5$, both $f_{res}$ and $f_{phase}$ are decreasing functions of $\tau_1$. In contrast to the $I_h + I_p$ model, in the $I_{Ks} + I_p$ model, $f_{res}$ and $f_{phase}$ are larger for $G_p > 0$ than for $G_p = 0$, although the difference between $f_{res}$ for these two values of $G_p$ is very small.

The two models show also similarities and differences in their natural frequencies as a function of $\tau_1$. As mentioned above, neither model shows intrinsic oscillations for $G_p = 0$. For $G_p > 0$, in both models $f_{res}$ is always larger than $f_{nat}$ (Figs. 11A1 and 11B1). For the $I_h + I_p$ model (Fig. 11A1), $f_{res}$ and $f_{nat}$ are almost identical for all values of $\tau_1$ (and $G_p > 0$), while the value of $f_{res}$ is significantly different than for $G_p = 0$. In contrast, for the $I_{Ks} + I_p$ model (Fig. 11B1), intrinsic oscillations (for $G_p > 0$) disappear as $\tau_1$ increases and $f_{res}$ is not significantly different for different values of $G_p$.

Resonance is amplified as $\tau_1$ increases in the presence of $I_p$ in the two models. In the absence of $I_p$ ($G_p = 0$) both $Z_{max}$ and $Q_Z$ are insensitive to changes in the time constant $\tau_1$ (Figs. 11A2 and 11B2). However, for $G_p = 0.5$, both $Z_{max}$ and $Q_Z$ increase as $\tau_1$ increases. Their difference remains constant since changes in $\tau_1$ have no effect on $Z(0)$. Note that changes in $Z_{max}$ and $Q_Z$ with $\tau_1$ are more pronounced for the $I_h + I_p$ model (Fig. 11A2) than for the $I_{Ks} + I_p$ model (Fig. 11B2).

Finally, although for the sake of brevity we do not show these data, $f_{res}$ and $f_{phase}$ are almost insensitive to changes in $I_{app}$ in the two models whereas resonance ($Z_{max}$ and $Q_Z$) is significantly amplified by increasing values of $I_{app}$ in the presence of $I_p$ in both models.

## Discussion

A large number of studies in recent years has focused on subthreshold membrane resonance in neurons, its dependence on ionic currents and its influence on neuronal and network oscillations (Castro-Alamancos et al. 2007; Ledoux and N. 2011; Richardson et al. 2003; Tohidi and Nadim 2009). Subthreshold membrane resonance is primarily described on the basis of a linear *RLC* circuit where *R* is equated with the membrane resistance, *C* with the membrane capacitance and the inductance *L*, more abstractly, with voltage-gated ionic conductance properties (Erchova et al. 2004). Membrane resonance is defined by a peak in the impedance amplitude plotted as a function of input frequency (the impedance profile) and can be characterized by a few attributes describing this impedance profile (Fig. 1B1). The frequency and quality ("peakiness") of membrane resonance is captured by these attributes. A number of voltage-gated ionic currents are known to result in membrane resonance. Yet, even in a simple model neuron that is based on biophysical currents, the dependence of the attributes of the impedance profile on biophysical parameters such as the maximal conductances of the leak and voltage-gated currents is neither well-described nor intuitive.

In the first part of this paper we used a linear 2D model (Richardson et al. 2003) to show how the attributes of the resonance and phase profiles depend on three model parameters, the two effective conductances and the time constant. This analysis resulted in graphs that show how resonance and phase depend on the model parameters (Figs. 2-5). We also compared the resonance frequency in this model with the frequency of subthreshold oscillations, when present (Fig. 4). In the second part, we used this description to understand the emergence and properties of resonance and phase-frequency relationships in two representative conductance-based biophysical models. Our approach was to explore first how changing a biophysical parameter changed the linearization about an equilibrium point of the conductance-based model. We then mapped the resulting trajectories (parameterized by the biophysical parameter) in the effective-conductance parameter space described in part one. These trajectories provided us with information on the dependence of resonance and phase attributes on the biophysical parameter. Together, these two parts provide a framework to study linear impedance and to analyze how changing biophysical parameters affects linear resonance and phase profiles in nonlinear conductance-based models. This framework was described for 2D systems but can be readily generalized to larger models.

We identified two qualitatively different mechanisms for the generation of resonance in 2D systems. The first mechanism depends on the effective conductance $\gamma_1$: as $\gamma_1$ increases, there is a decrease in both $Z_{max}$ and $Z(0)$ but the latter decreases at a higher rate (Fig. 3C) which results in an increase in $Q_Z$. The second mechanism is due to changing $\tau_1$. For very small values of $\tau_1$ the 2D system does not exhibit resonance (Fig. 6A). As $\tau_1$ increases, the system first transitions through a region of the parameter space which shows intrinsic oscillations with no resonance and for larger values of $\tau_1$ resonance emerges (Fig. 6A-B). Increasing $\tau_1$ further can allow the system to resonance without subthreshold oscillations (Fig 6B1-B2). As such, factors such as temperature that modulate time constants can allow a neuron to transition to resonance without changing the levels of expressed ionic currents.

The phenomenon of subthreshold resonance is often related to the presence of subthreshold oscillations (Lampl and Yarom 1997) although it is known that these are separate phenomena even in linear models (Richardson et al. 2003). The analysis of the linear 2D model shows that in certain parameter ranges these two frequencies can show similar parameter dependence but that they are also clearly distinct (Fig. 4).

Typically the voltage response of a neuron is studied by focusing on the amplitude of the impedance profile (Hutcheon and Yarom 2000). Yet, just as the impedance amplitude describes the scaling of the membrane potential, the impedance phase as a function of frequency (the phase profile) describes the membrane potential phase shift in response to an oscillatory current. The zero-phase frequency $f_{phase}$—the frequency at which the input current and the membrane potential are synchronous in phase—provides information complementary to the resonance frequency $f_{max}$ (Richardson et al. 2003). Although in an *RLC* circuit in series the values of $f_{phase}$ and $f_{max}$ are equal, in parallel circuits, such as membrane equivalent circuits, the two values remain distinct (Fig. 1B). Here we examined the dependence of $f_{phase}$ in a 2D system and found that this frequency remains independent of the effective leak conductance, although the extent to which negative phase shifts occur ($\phi_{min}$) does change with both effective leak and voltage-gated conductances (Fig. 5).

Voltage-gated ionic conductances have been classified into resonant and amplifying (Hutcheon and Yarom 2000; Richardson et al. 2003) with the underlying implication that all currents in each group behave similarly with respect to changes in parameters. Here we explored two representative conductance-based models with distinct resonant currents, one hyperpolarization-activated ($I_h$) and the other depolarization-activated ($I_{Ks}$). We found that, in the absence of an amplifying current ($I_p$), the resonant attributes in both models exhibit a similar dependence on the conductance of the two respective resonant currents. This is most easily seen by comparing the solid trajectories in Figs. 8A, B1 and C1 with the corresponding panels of Fig. 9. In contrast, in the presence of the amplifying current $I_p$, changing the conductance of the resonant current in the two models has opposite effects on the attributes of resonance. For instance, whereas increasing $G_h$ can amplify the voltage response, a similar increase in $G_q$ does the opposite (Figs. 8B2 and 9B2). Similarly, increasing these two conductances in the respective models can have opposite effects on the properties of the phase profile (Figs. 8C2 and 9C2). The difference in the effects of $G_h$ and $G_q$ are again readily observed by examining the corresponding trajectories in the effective-conductance parameter space (compare dashed curves in Figs. 8A and 9A). Although these differences are expected, they have been previously overlooked in the analysis of membrane resonance in neural models.

In this study we have explored the properties of resonance and phase profiles in a linear system and used this information to examine similar properties for linearizations of simple conductance-based models. Interestingly, some effects of the nonlinearities in the models are captured by the nonlinear trajectories in the effective-conductance parameter space. Yet, it is important to note that the linearizations do not capture other effects that may arise due to nonlinearities that are present even in these simplified conductance-based models. Such effects include, but are not limited to, nonlinear amplifications of the voltage response with input current and the emergence of voltage responses that do not match the input frequency. We will address these issues in future work.

# Appendix A. Two-dimensional linear systems: eigenvalues, natural frequency and impedance

Consider the following two-dimensional linear system

$$\begin{cases} X' = aX + bY + A_{in} e^{i\omega t} \\ Y' = cX + dY \end{cases} \quad (17)$$

$a$, $b$, $c$ and $d$ are constant, $\omega > 0$ and $A_{in} \geq 0$.

*Intrinsic oscillations and natural frequency*

The Jacobian of the corresponding homogeneous system ($A_{in} = 0$) is given by

$$J = \begin{pmatrix} a & b \\ c & d \end{pmatrix}.$$

The roots of the characteristic polynomial are given by

$$r_{1,2} = \frac{(a+d) \pm \sqrt{(a-d)^2 + 4bc}}{2}. \quad (18)$$

From Eq. (18), the homogeneous system displays stable oscillatory solutions for values of the parameters satisfying

$$(a-d)^2 + 4bc < 0$$
$$a + d \leq 0.$$

The natural frequency is given by

$$\mu = \frac{\sqrt{4bc + (a-d)^2}}{2}. \quad (19)$$

*Impedance function*

The particular solutions to system (17) have the form

$$X_p(t) = A_{out} e^{i\omega t}$$
$$Y_p(t) = B_{out} e^{i\omega t}$$

Substituting into system (17) and rearranging terms we obtain

$$\begin{pmatrix} (i\omega - a) & -b \\ -c & (i\omega - d) \end{pmatrix} \begin{pmatrix} A_{out} \\ B_{out} \end{pmatrix} = \begin{pmatrix} A_{in} \\ 0 \end{pmatrix}. \quad (20)$$

By solving the algebraic system (20) we obtain

$$Z(\omega) = \frac{A_{out}}{A_{in}} = \frac{-d + i\omega}{(ad - bc - \omega^2) - i\omega(a+d)},$$

$$|Z(\omega)|^2 = \frac{A_{out}^2}{A_{in}^2} = \frac{d^2 + \omega^2}{(ad - bc - \omega^2)^2 + (a+d)^2 \omega^2}. \quad (21)$$

The impedance $Z$ peaks at the resonant frequency

$$\omega_{res} = \sqrt{-d^2 + \sqrt{b^2c^2 - 2abcd - 2d^2bc}}. \tag{22}$$

For the calculation of the phase, we note that

$$C_1 = -\frac{(ad-bc-\omega^2)d + (a+d)\omega^2}{(ad-bc-\omega^2)^2 + (a+d)^2\omega^2}$$

$$C_2 = \frac{(ad-bc-\omega^2)\omega - (a+d)\omega d}{(ad-bc-\omega^2)^2 + (a+d)^2\omega^2}$$

Then

$$\tan(\phi) = \frac{(ad-bc-\omega^2)\omega - (a+d)\omega d}{(ad-bc-\omega^2)d + (a+d)\omega^2} = -\frac{bc+d^2+\omega^2}{(ad-bc)d + a\omega^2}\omega. \tag{23}$$

## Appendix B. The 2D Biophysical Models

The voltage-dependent activation/inactivation curves $x_\infty(V)$ and voltage-dependent time scales $\tau_x(V)$ used in the kinetic Eq. (3) for the gating variables $x$ are frequently expressed in terms of

$$x_\infty(V) = \frac{\alpha_x(V)}{\alpha_x(V) + \beta_x(V)}$$

$$\tau_x(V) = \frac{1}{\alpha_x(V) + \beta_x(V)}. \tag{24}$$

Below we present the functions and parameters corresponding to the various models we use in this paper.

### $I_h + I_p$ model

This model has been proposed in (Acker et al. 2003). It has a persistent sodium current and a two-component (fast and slow) h-current given by $I_p = G_p p(V - E_{Na}) = G_p p_\infty(V)(V - E_{Na})$ and $I_h = G_h r(V - E_h) = G_h(c_f r_f + c_s r_s)(V - E_h)$ respectively with $E_h = -20$ mV, $E_{Na} = 55$ mV, $c_f = 0.65$ and $c_s = 0.35$. The voltage-dependent activation/inactivation and time constants are given by

$$p_\infty(V) = 1/(1+e^{-(V+38)/6.5})$$
$$\tau_p(V) = 0.15$$
$$r_{f,\infty}(V) = 1/(1+e^{(V+79.2)/9.78})$$
$$\tau_{r_f}(V) = 0.51/(e^{(V-1.7)/10} + e^{-(V+340)/52}) + 1$$
$$r_{s,\infty} = 1/(1+e^{(V+71.3)/7.9})$$
$$\tau_{r_s}(V) = 5.6/(e^{(V-1.7)/14} + e^{-(V+260)/43}) + 1.$$

## $I_{Ks} + I_p$ model

This model has been proposed in (Acker et al. 2003). It has a persistent sodium current and a slow potassium (M-type) current given by $I_p = G_p p(V - E_{Na}) = G_p p_\infty(V)(V - E_{Na})$ and $I_{Ks} = G_q q(V - E_K)$ with $E_{Na} = 55$ mV and $E_K = -90$ mV. The voltage-dependent activation/inactivation and time constants are given by

$$p_\infty(V) = 1/(1+e^{-(V+38)/6.5})$$
$$\tau_p(V) = 0.15$$
$$q_\infty(V) = 1/(1+e^{-(V+35)/6.5})$$
$$q_\tau(V) = 90$$

## Acknowledgements

The authors thank Diana Martinez and David Fox for their comments on this manuscript.

## Grants

Supported by NSF DMS0817241 (HGR) and NIH MH060605 (FN).

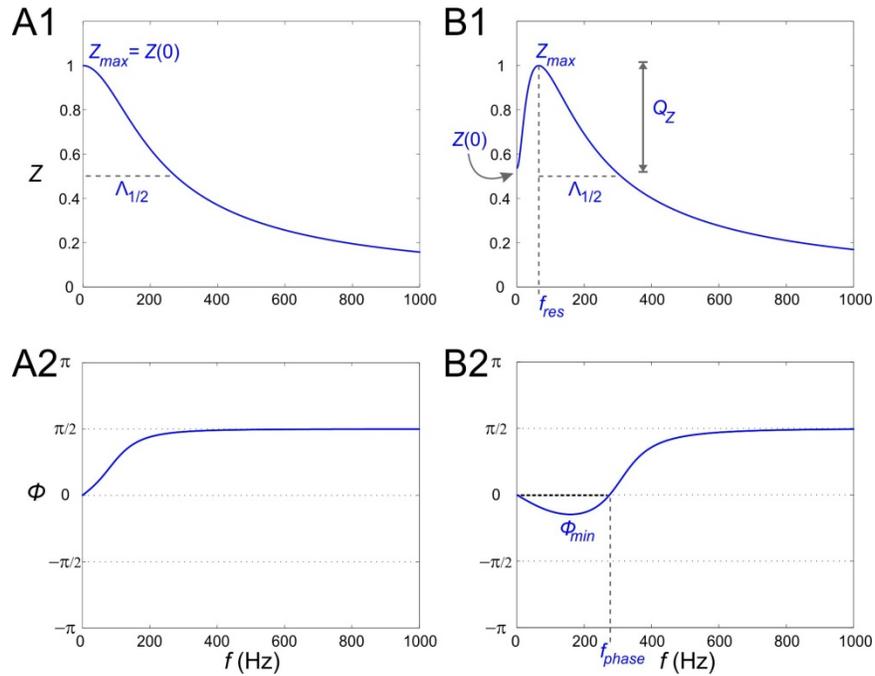

**Figure 1.** Impedance ($Z$) and phase ($\phi$) profile diagrams. **A1.** The impedance profile of a low-pass filter RC cell. A2. The RC cell in A1 only shows a positive response phase. **B1.** The impedance profile of a band-pass filter resonant cell. phase only). The resonant frequency $f_{res}$ is the frequency at which the impedance function reaches its maximum $Z_{max}$. The parameter $Q_Z$ is defined as the difference between $Z_{max}$ and $Z_0$. The characteristic frequency range $\Lambda_{1/2}$ measures the length of the frequency interval for which $Z(f)$ is above ½ of its maximum value. **B2.** The phase of the resonant cell shown in B1 includes both a negative and a positive region. $\Phi_{min}$ is the minimum phase and $f_{phase}$ represents the zero-phase frequency and is equal to the length of the frequency band for which the voltage response leads the input signal (negative phase).

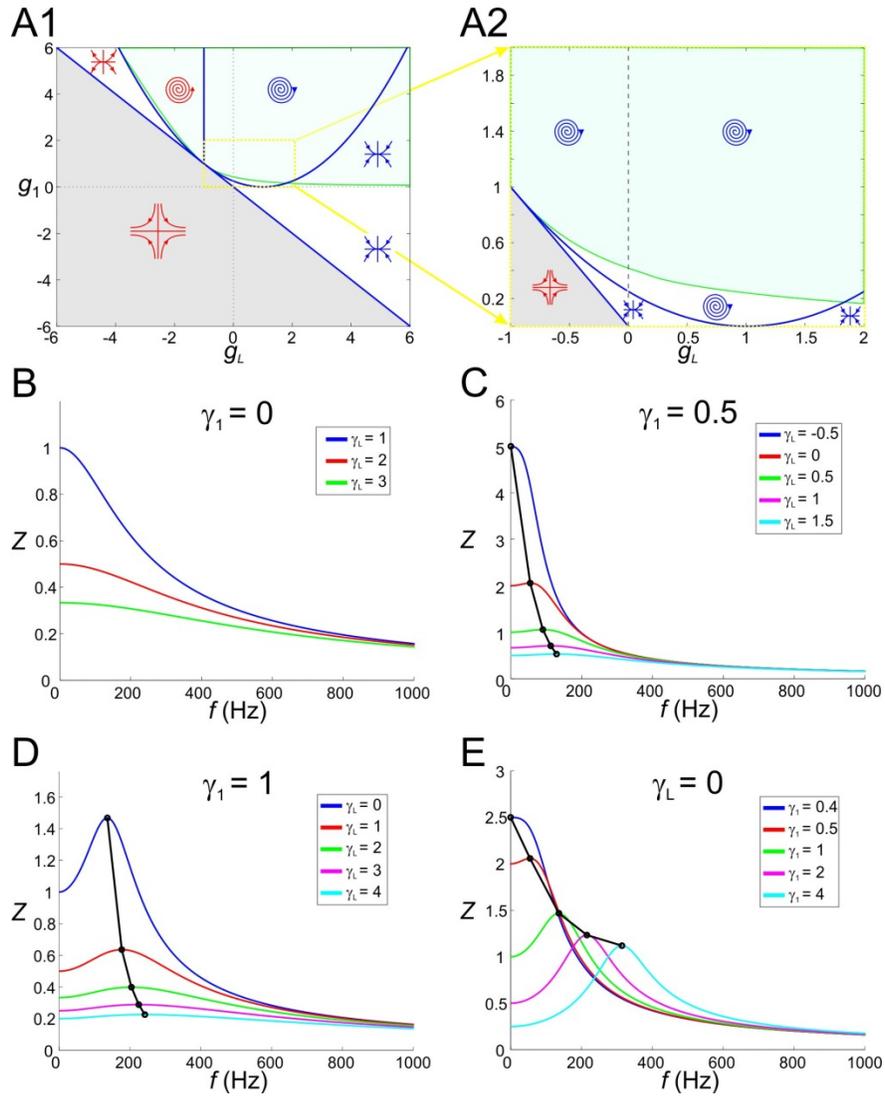

**Figure 2.** Stability, oscillatory, and resonant properties for the 2D linear system **(6)** With $C = 1$ and $\tau_1 = 1$ in the $g_L$ - $g_1$ space. (For these parameters values, $g_1 = \gamma_1$ and $g_L = \gamma_L$ and the 2D linear **(6)** and the 2D rescaled system **(10)** are equivalent.) **A1.** Superimposed stability and resonance diagrams. The blue curves separate between regions in parameter space with different stability properties or fixed point types. Symbols indicate whether the fixed point is a focus, node or saddle (blue symbols stable; red symbols unstable). The green curve (resonance curve) separates between regions in parameter space where the system exhibits resonance (green) or does not (other). **A2.** A magnification of A1 (yellow dotted box). All regions above the grey saddle region are stable. **B. to D.** Representative examples of impedance profiles for the rescaled system **(10)** with $\gamma_1$ and $\gamma_L$ values denoted in the figure. Black dots and lines denote the resonant points.

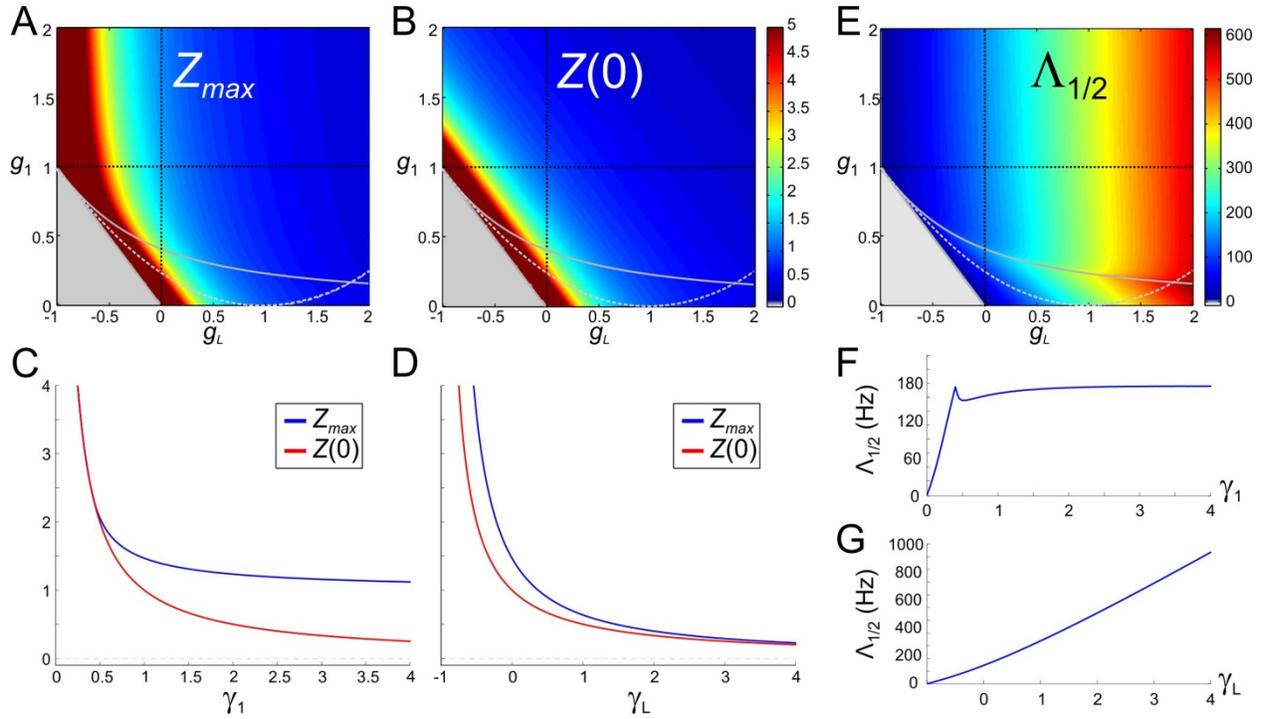

**Figure 3.** The parameters defining resonance for the 2D linear system **(6)** with $C = 1$ and $\tau_1 = 1$ in the $g_L$-$g_1$ space. (For these parameters values, $g_1 = \gamma_1$ and $g_L = \gamma_L$ and the 2D linear **(6)** and the 2D rescaled system **(10)** are equivalent.) **A.** The maximum impedance $Z_{max}$. **B.** The zero-frequency input resistance $Z(0)$. The scale bar in B also denotes the scale in A. The red regions in A and B are outside the scale of the graph. **C.** Representative example of $Z_{max}$ and $Z(0)$ as a function of $\gamma_1$ for $\gamma_L = 0$ (vertical dashed black line in A and B). **D.** Representative example of $Z_{max}$ and $Z(0)$ as a function of $\gamma_L$ for $\gamma_1 = 1$ (horizontal dashed black line in A and B). **E.** Half band-width $\Lambda_{1/2}$. **F.** Representative example of $\Lambda_{1/2}$ as a function of $\gamma_1$ for $\gamma_L = 0$ (vertical dashed black line in E). **G.** Representative example of $\Lambda_{1/2}$ as a function of $\gamma_L$ for $\gamma_1 = 1$ (horizontal dashed black line in E).

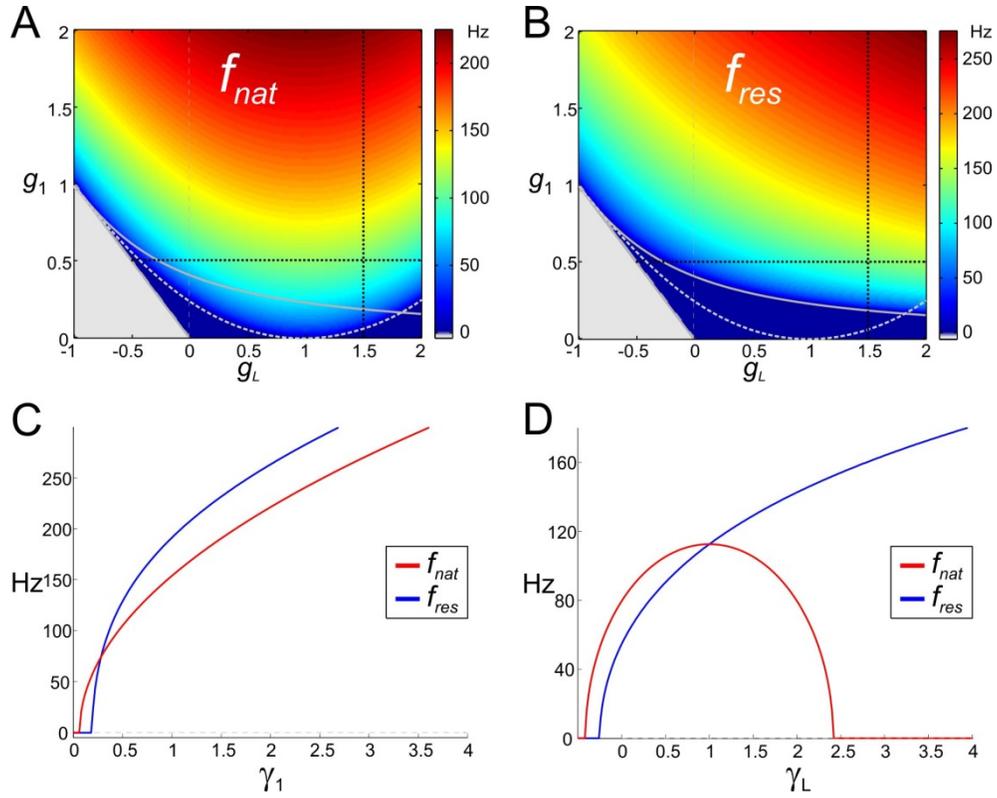

**Figure 4.** Natural and resonant frequency color bar diagrams for the 2D linear system **(6)** with $C=1$ and $\tau_1=1$ in the $g_L$-$g_1$ space. (For these parameters values, $g_1 = \gamma_1$ and $g_L = \gamma_L$ and the 2D linear **(6)** and the 2D rescaled system **(10)** are equivalent.) **A.** Natural frequency $f_{nat}$. **B.** Resonant frequency $f_{res}$. The dashed and solid gray curves in panels B and C are the stability (blue) and resonance (red) curves in panel A. The shaded gray region in parameter space below the line $g_1 = -g_L$ in panels B and C is a region of saddle points. **C.** Representative example of $f_{res}$ and $f_{nat}$ as a function of $\gamma_1$ for $\gamma_L = 1.5$ (vertical dashed black lines in A and B). **D.** Representative example of $f_{res}$ and $f_{nat}$ as a function of $\gamma_L$ for $\gamma_1 = 0.5$ (horizontal dashed black lines in A and B).

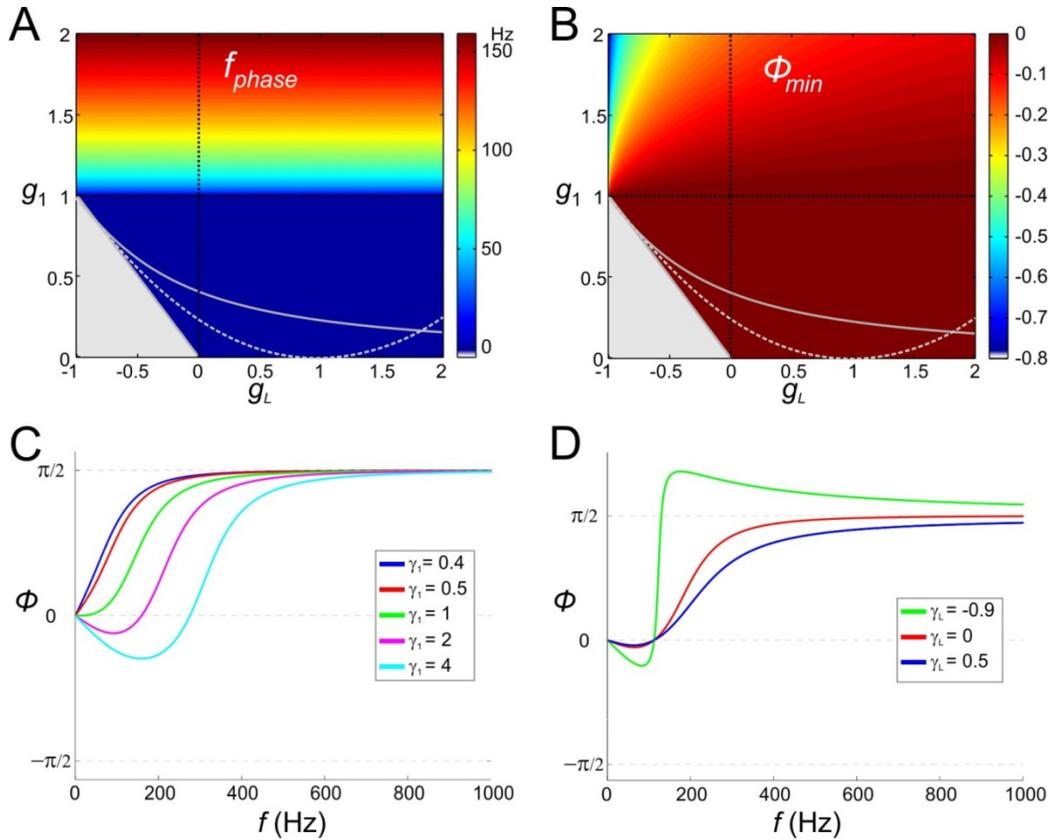

**Figure 5.** Negative-phase frequency-band for the 2D linear system **(6)** with $C = 1$ and $\tau_1 = 1$ in the $g_L$ - $g_1$ space. (For these parameters values, $g_1 = \gamma_1$ and $g_L = \gamma_L$ and the 2D linear **(6)** and the 2D rescaled system **(10)** are equivalent.) **A.** The length of the negative-phase frequency band (or, equivalently, the zero-phase frequency $f_{phase}$). **B.** Minimum phase ($\phi_{min}$) of the voltage response. **C.** Representative example of the phase profile as a function of $\gamma_1$ for $\gamma_L = 0$ (vertical dashed black line in A and B). **D.** Representative example of the phase profile as a function of $\gamma_L$ for $\gamma_1 = 1$ (vertical dashed black line in A and B).

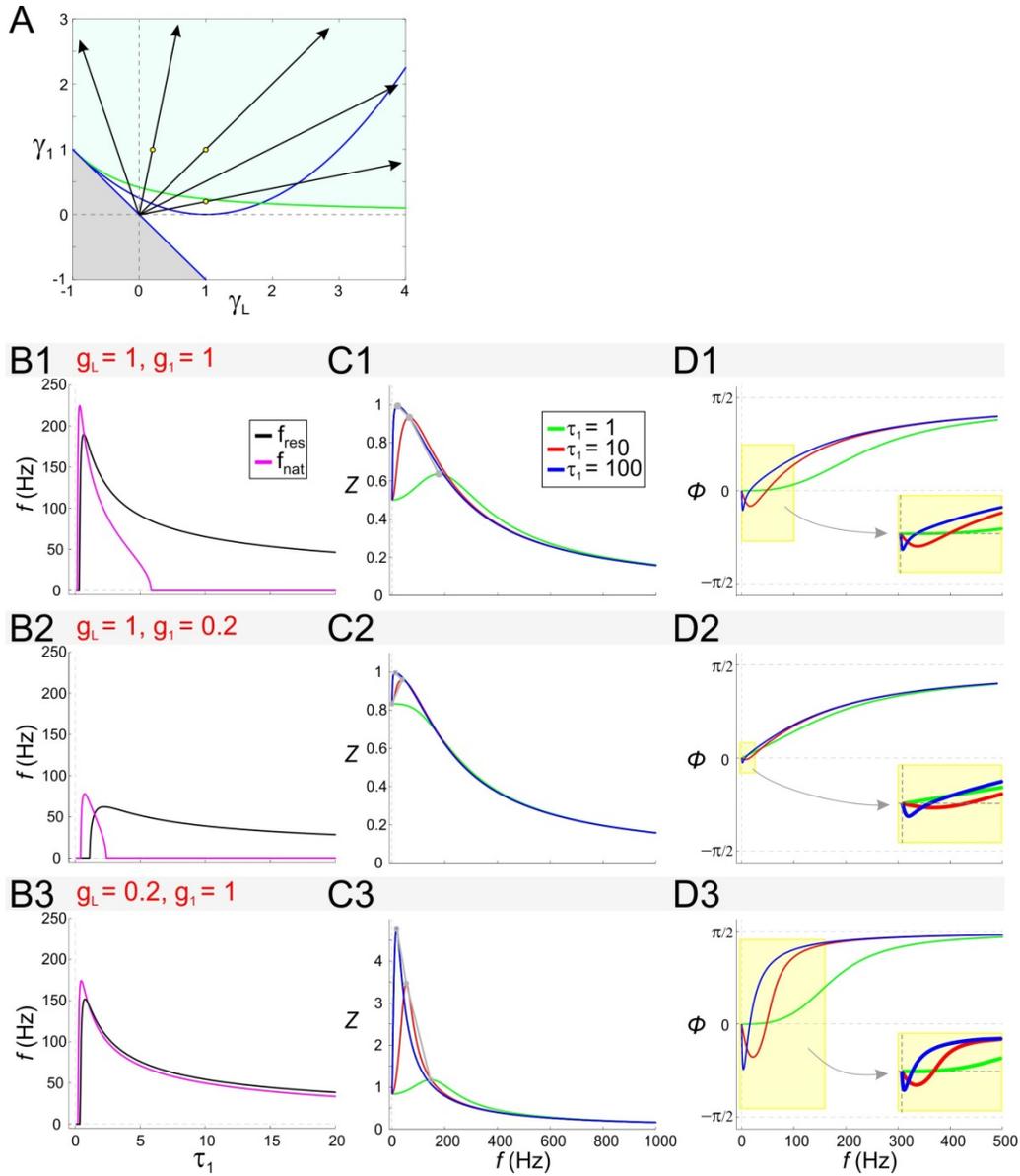

**Figure 6.** The effect of the time constant $\tau_1$ on impedance and phase in the linear model. **A.** The effect of $\tau_1$ on the rescaled conductances $\gamma_L$ and $\gamma_1$. The region shown is as in Fig. 2A. For fixed values of the conductances $g_1$ and $g_L$, as $\tau_1$ increases, the rescaled effective conductances $\gamma_1$ and $\gamma_L$ increase along lines emanating from the origin. **B.** Resonant and natural frequencies of the 2D linear system as a function of $\tau_1$ for representative values of $g_1$ and $g_L$. **C.** Impedance profiles for representative values of $g_1$ and $g_L$ with different values of $\tau_1$ (as shown in C1). **D.** Phase profiles for representative values of $g_1$ and $g_L$ with different values of $\tau_1$ (as shown in C1). Insets show expansions of the shaded box and demonstrate that increasing $\tau_1$ moves the zero-phase frequency to smaller values.

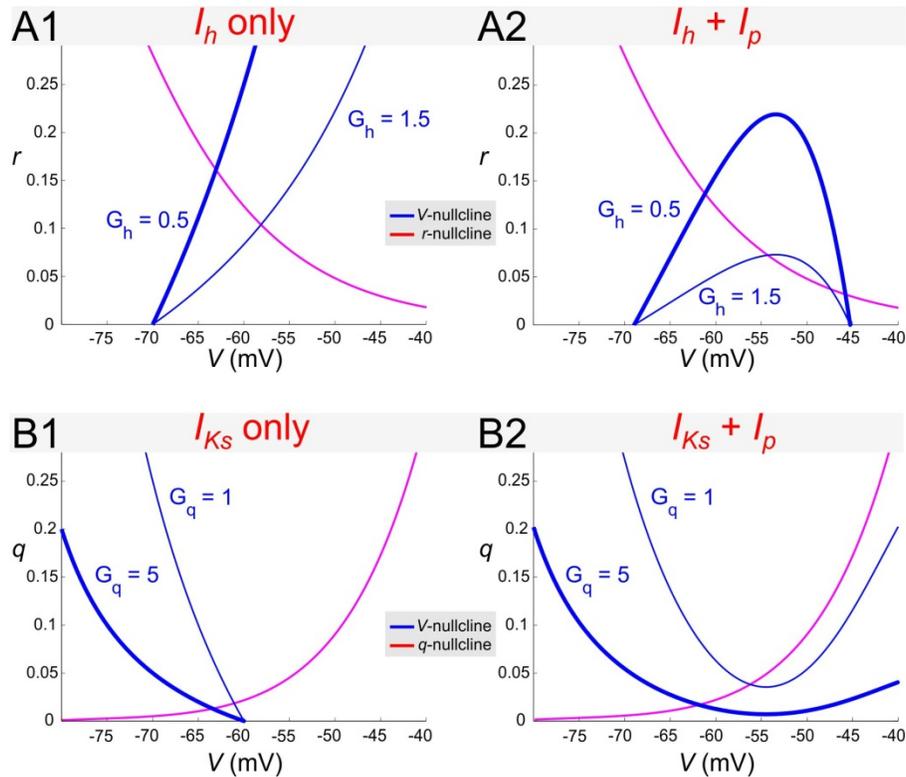

**Figure 7.** Nullclines for different 2D biophysical models. **A.** In the $I_h + I_p$ model with $G_p = 0$ (**A1**) different values of $G_h$ lead to nullclines that are close to linear near the fixed point. In contrast, when $G_p > 0$ (**A2**; $G_p = 0.5$) the V-nullcline can show quadratic nonlinearities near the fixed point. **B.** Similarly in the $I_M + I_p$ model with $G_p = 0$ (**B1**) the nullclines are close to linear whereas when $G_p > 0$ (**B2**; $G_p = 0.5$) the V-nullcline shows quadratic nonlinearities. Note that the nullclines are qualitatively mirror images of those of the $I_h + I_p$ model.

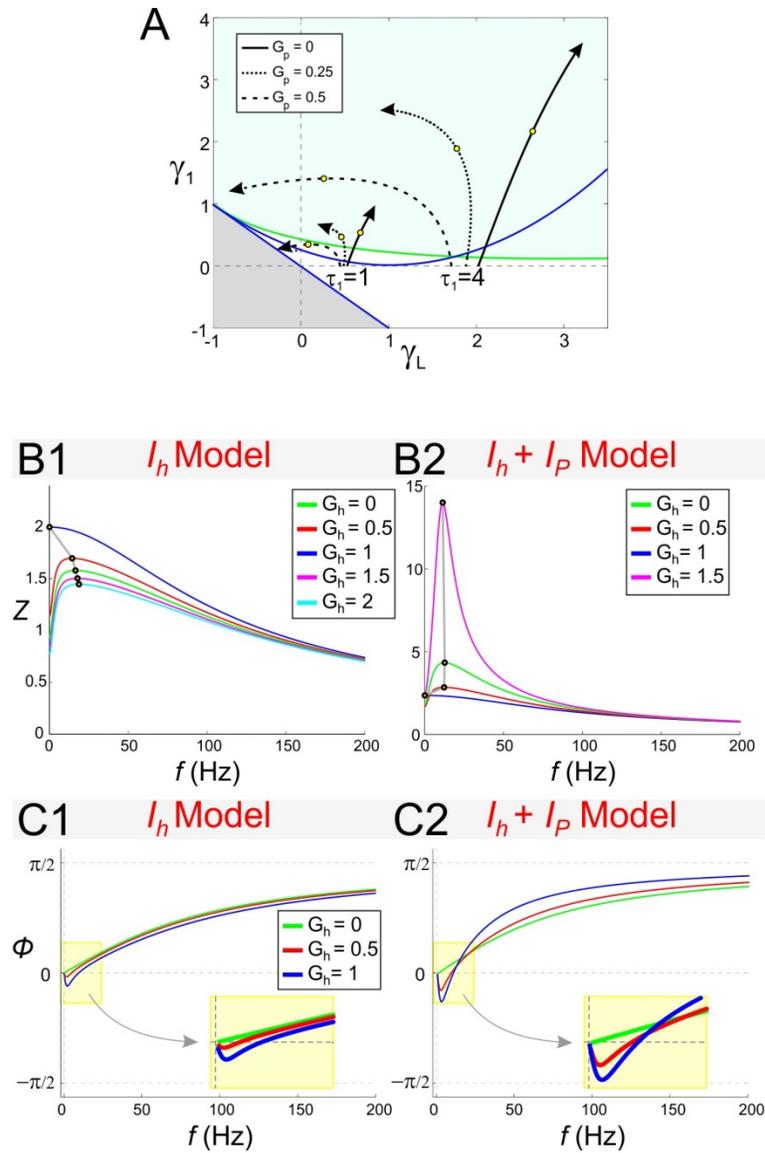

**Figure 8.** The effect of $G_h$ on resonance and zero-phase frequency in the $I_h + I_p$ model. **A.** Trajectories in $\gamma_L$-$\gamma_1$ space parameterized by the resonant conductance $G_h$ (arrows indicate increasing values of $G_h$). All trajectories start at $G_h = 0$ and the value $G_h = 1.5$ is marked by an open circle. Different trajectories correspond to different representative values of $G_p$ and $\tau_1$ (as indicated). The trajectories are computed until the fixed-point $\bar{V}$ ceases to exist. **B.** Impedance profiles for the linearized $I_h + I_p$ model with $G_p = 0$ (**B1**) and $G_p = 0.5$ (**B2**). **C.** Phase profiles for the linearized $I_h + I_p$ model with $G_p = 0$ (**C1**) and $G_p = 0.5$ (**C2**). In **B** and **C** $I_{app} = -2.5$.

Page | 32

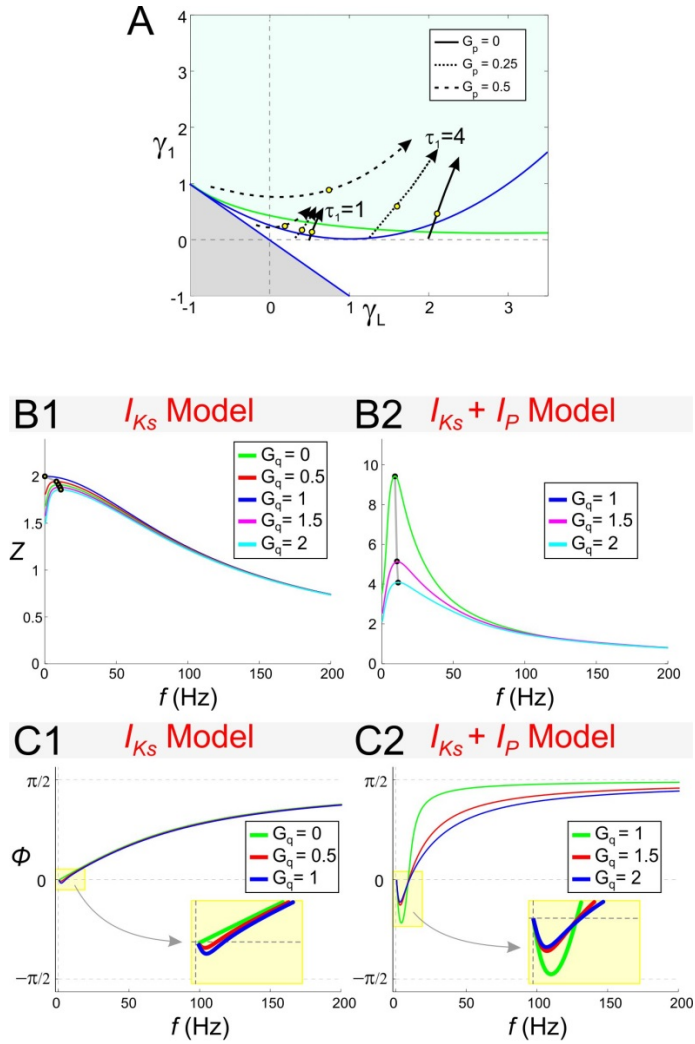

**Figure 9.** The effect of $G_q$ on resonance and zero-phase frequency in the $I_{Ks} + I_p$ model. **A.** Trajectories in $\gamma_L$-$\gamma_1$ space parameterized by the resonant conductance $G_q$ (arrows indicate decreasing values of $G_q$). Each trajectory starts at the value of $G_q$ at which the two nullclines first intersect and $\bar{V}$ emerges. The end of each trajectory marks $G_q = 3$ in all cases and open circles mark $G_q = 1$. Different trajectories correspond to different representative values of $G_p$ and $\tau_1$ (as indicated). **B.** Impedance profiles for the linearized $I_{Ks} + I_p$ model with $G_p = 0$ (**B1**) and $G_p = 0.5$ (**B2**). **C.** Phase profiles for the linearized $I_{Ks} + I_p$ model with $G_p = 0$ (**C1**) and $G_p = 0.5$ (**C2**). In **B** and **C** $I_{app} = 2.5$.

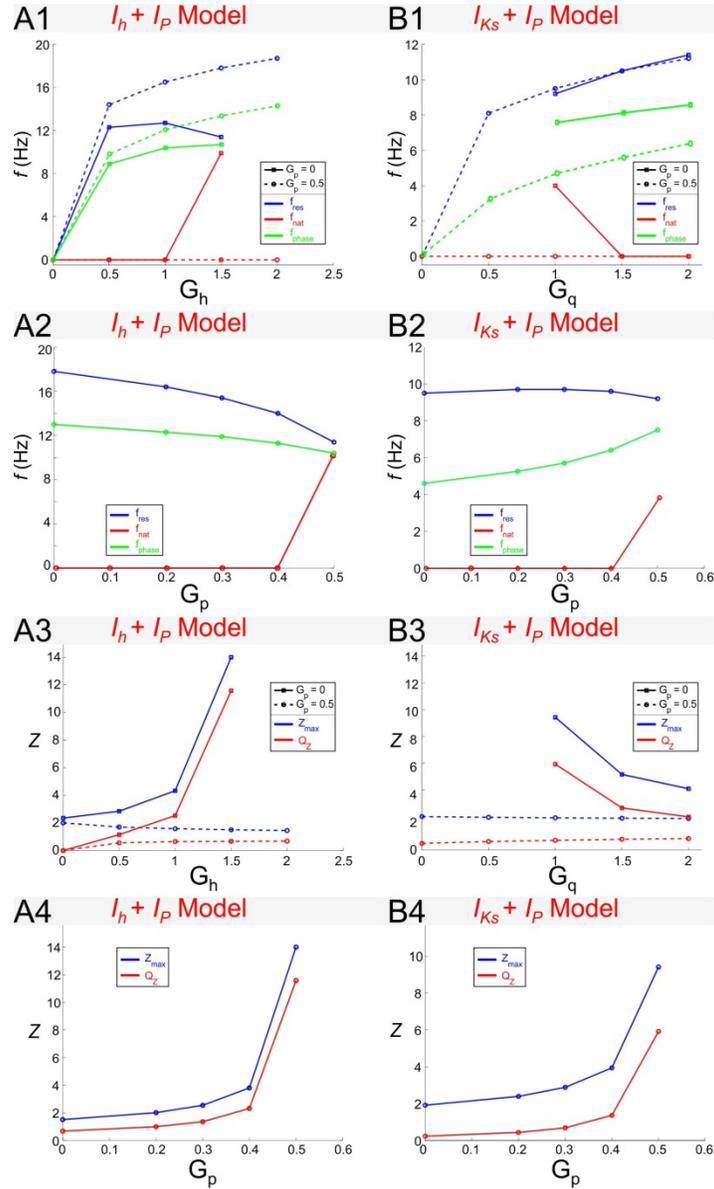

**Figure 10.** Dependence of resonance and phase properties for the $I_h + I_p$ (column **A**) and $I_{Ks} + I_p$ (column **B**) models on the maximal conductances. Panels **A1** and **B1** show $f_{res}$, $f_{nat}$ and $f_{phase}$ as a function of $G_h$ (**A1**) or $G_q$ (**B1**) with $G_p = 0$ or $G_p = 0.5$. Panels **A2** and **B2** show $f_{res}$, $f_{nat}$ and $f_{phase}$ as a function of $G_p$. Panels **A3** and **B3** show $Z_{max}$ and $Q_Z$ as a function of $G_h$ (**A3**) or $G_q$ (**B3**). $f_{res}$ and $f_{nat}$ as a function of $G_q$. Panels **A4** and **B4** show $Z_{max}$ and $Q_Z$ as a function of $G_p$. $I_{app} = -2.5$ for the $I_h + I_p$ model and $I_{app} = 2.5$ for the $I_{Ks} + I_p$ model. The time constant $\tau_1 = 80$ for both models.

Page | 34

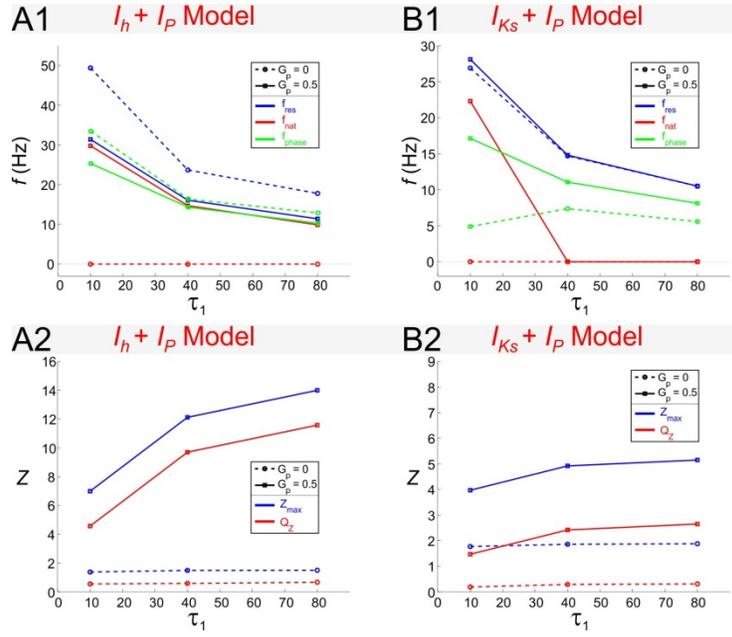

**Figure 11.** Dependence of resonance and phase properties for the $I_h + I_p$ (column **A**) and $I_{Ks} + I_p$ (column **B**) models on the time constant $\tau_1$. Panels **A1** and **B1** show $f_{res}$, $f_{nat}$ and $f_{phase}$ as a function of $\tau_1$. Panels **A2** and **B2** show $Z_{max}$ and $Q_Z$ as a function of $\tau_1$. $I_{app} = -2.5$ for the $I_h + I_p$ model and $I_{app} = 2.5$ for the $I_{Ks} + I_p$ model. The time constant $\tau_1 = 80$ for both models.